\documentclass[eqsecnum,aps,article]{revtex4}
\usepackage{graphicx}
\usepackage{amsmath}
\usepackage{amsfonts}
\usepackage{color}

\begin{document}
\author{S. Chaturvedi\footnote{email: scsp@uohyd.ernet.in}}
\address{School of Physics, University of Hyberabad, Hyberabad 500 046, India}
\author{E. Ercolessi\footnote{email: ercolessi@bo.infn.it}}
\address{Dipartimento di Fisica, Universit\`a di Bologna and INFN, Via Irnerio 46, 40126 Bologna, Italy}
\author{A. Ibort \footnote{On leave of absence from Departamento de Matem\'aticas, Universidad Carlos III de Madrid, Spain}\footnote{email: albertoi@math.uc3m.es}}
\address{Department of Mathematics, Univ. of California at Berkeley, Berkeley CA 94720, USA} 
\author{G. Marmo\footnote{email: marmo@na.infn.it}}
\address{Dipartimento di Scienze Fisiche, Universit\`a di Napoli Federico II and INFN, Via Cinzia, 80126 Napoli, Italy}
\author{G. Morandi\footnote{email: morandi@bo.infn.it}}
\address{Dipartimento di Fisica, Universit\`a di Bologna and INFN, Via Irnerio 46, 40126 Bologna, Italy} 
\author{N. Mukunda\footnote{email: nmukunda@gmail.com}}
\address{The Institute of Mathematical Sciences, C.I.T. Campus, Tharamani, Chennai 600 113, India}
\author{R. Simon\footnote{email: simon@imsc.res.in}}
\address{The Institute of Mathematical Sciences, C.I.T. Campus, Tharamani, Chennai 600 113, India}

\title{Null Phase Curves and Manifolds in Geometric Phase Theory}
\begin{abstract}
Bargmann invariants and null phase curves are known to be important
ingredients in understanding the essential nature of the geometric phase in
quantum mechanics. Null phase manifolds in quantum-mechanical ray spaces are
submanifolds made up entirely of null phase curves, and so are equally
important for geometric phase considerations. It is shown that the complete
characterization of null phase manifolds involves both the Riemannian metric
structure and the symplectic structure of ray space in equal measure, which
thus brings together these two aspects in a natural manner.
\end{abstract}
 \maketitle

\section{Introduction} \label{s:introd}

The understanding of the structure and properties of the geometric phase in
quantum mechanics, originally discovered in the context of unitary adiabatic
cyclic Schr\"{o}dinger evolution \cite{BE}, have improved considerably on account of
several important later developments. Thus it became clear in successive
stages that neither the adiabatic condition nor the cyclic condition are
necessary for the existence and identification of the geometric phase \cite{SI,AA}. In
the latter step, an important role was played by the exploitation of the
fact that the state space describing the pure states of a quantum system
carries a Riemannian metric, leading to corresponding geodesics in this
space. These geodesics were used to convert a general non-cyclic quantum
evolution to a cyclic one, so that previous definitions of the geometric
phase could then be used to show its existence. The third significant step
was the elucidation of a purely kinematical approach to the geometric phase
in which the Schr\"{o}dinger equation and a hermitian hamiltonian operator
were both shown to be inessential \cite{SB}

Several precursors to the quantum-mechanical geometric phase concept have
been recognized. Of these, it may be argued that the work of Pancharatnam \cite{MS1}
in the context of interference phenomena in classical polarization optics,
and of Bargmann in the context of the Wigner unitary-antiunitary theorem
for symmetry operations in quantum mechanics \cite{PA}, are particularly significant.
Pancharatnam's work has led to the fruitful concept of two
quantum-mechanical Hilbert space vectors being in phase with respect to one
another, and more generally to a measure of their relative phase. The
phase found by him in polarization optics has been seen later to be an early
manifestation of the geometric phase in a decidedly non-adiabatic though
cyclic situation.

Bargmann's work introduced a family of complex expressions into quantum
bechanics, later given the name ``Bargmann invariants", which capture in
powerful and elegant terms the essential role of complex numbers in the
mathematical formalism of quantum mechanics. One of the outcomes of the
kinematical approach to geometric phases has been to bring out the
importance of the Bargmann invariants, and another has been to combine them
with the geodesics mentioned earlier to show that their phases are actually
geometric phases for certain cyclic evolutions \cite{SB}.

The deep interrelations that exist among the ideas of Pancharatnam, Bargmann and Berry
have been described elsewhere \cite{MUK}.

More recently, further exploration of the kinematical treatment of geometric
phases has led to the important concept of null phase curves (\text{NPC}) in quantum-mechanical Hilbert 
and ray spaces, which are a vast
generalization of geodesics but which preserve the connection between
Bargmann invariants and geometric phases \cite{RAM}. This work has shown that the
initial role of geodesics in geometric phase theory has been
essentially fortuitous, and that it is the far more numerous \text{NPC}'s that
really belong to this theory. Indeed, it has been shown that the entire
theory can \ be built up logically based on Bargmann invariants and \text{NPC}'s,
with the definition of the latter actually based on the former \cite{MAEMM}.

Traditional expositions of quantum mechanics \ have tended to lay stress on
the complex linear structure of Hilbert spaces, the non-commutativity of
Hermitian operators representing physical observables, and then drawing out
various consequences. In more recent times, with the emphasis given to the
study of ray spaces that describe pure quantum states in a one-to-one
manner, the rich mathematical structures that come automatically with these
spaces have received a great deal of attention \cite{EMM}. Thus from the familiar
complex inner products among Hilbert space vectors there emerge both a
Riemannian structure (mentioned above) with a non-degenerate metric on ray
space, and a symplectic structure (a classical-looking phase space
structure) on the same ray space. Quantum mechanical ray spaces are
simultaneously Riemannian manifolds and symplectic manifolds, and this fact
would naturally be expected to have important physical manifestations and
consequences. The results presented in this work point in that direction.

It has been mentioned that \text{NPC}'s are far more numerous than geodesics.
This is so to such an extent that it seems reasonable to ask if there are
submanifolds (of various dimensions) in quantum-mechanical ray spaces such
that every (sufficiently smooth) curve in any one of them is a \text{NPC}; and
if so, how such submanifolds can be characterized. Such submanifolds have
been called \textit{Null Phase Manifolds} (\text{NPM}) and examples
given \cite{MAEMM}. We take up their study here and will show that the characterization
of \text{NPM}'s indeed involves both the Riemannian structure (through its
geodesics) and the symplectic structure of ray space (through the concept of
isotropic submanifolds) in equal measure. It is quite remarkable that this
should be so, and it suggests that \text{NPM}'s are important for grasping the
mathematical structure of quantum mechanics at the deepest level.\\

The contents of this paper are arranged as follows.
Section $2$ collects basic notations relating to the Hilbert and ray spaces in
quantum mechanics, and the definition of Bargmann invariants and geometric
phases in the kinematic approach.
The role of ray space geodesics in providing a connection between Bargmann
invariants and geometric phases is sketched. After introducing the \text{NPC}
concept, the greatly enlarged nature of this connection is mentioned.
Section $3$ begins with a set of basic relations involving Geometric Phases,
\text{NPC}'s and the symplectic two-form on ray space. The general definition of a
\text{NPM} in ray space is then given. While it is easy to see that a \text{NPM} is
necessarily isotropic (with respect to the ray space symplectic structure),
the converse is not true.
It is then shown by explicit construction that the most general \text{NPM} can be
characterized as follows: it is a submanifold in an isotropic and totally
geodesic submanifold in ray space, though it may not itself be totally geodesic.
Section $4$ gives several examples of the construction of Sect. $3$, in
addition to a somewhat detailed description of a general \text{NPC}.
Section $5$ contains some concluding remarks.

\section{Bargmann Invariants, Geometric Phases and $\text{NPC}$'s} \label{s:Bargmann}

We begin by recalling basic notations and definitions from previous work. We
denote by $\mathcal{H}$ the complex Hilbert space pertaining to some quantum
system. Vectors and the inner product are denoted as $\psi ,\phi ,...$ and $%
\left( \phi ,\psi \right) $ respectively. The unit sphere $\mathcal{B}%
\subset \mathcal{H}$ and the ray space $\mathcal{R}$ are respectively:%
\begin{equation}
\begin{array}{c}
\mathcal{B}=\left\{ \psi \in \mathcal{H} \mid \text{ }\left( \psi ,\psi \right)
=1\right\} \subset \mathcal{H} ;\\ 
\mathcal{R}=\left\{ \rho \left( \psi \right) =\psi \psi ^{\dag } \mid \text{ }%
\psi \in \mathcal{B}\right\} 
\end{array}
\label{NPC.2.1}
\end{equation}

The projection: $\pi :\mathcal{B}\rightarrow \mathcal{R}$ maps $\psi $ to $%
\pi \left( \psi \right) =\rho \left( \psi \right) $, and \ $\mathcal{B}$ is
a $U\left( 1\right) $ principal bundle over $\mathcal{R}$.  If $\mathcal{H}$ is of
finite complex dimension $N$, the real dimension of $\mathcal{B}\simeq
S^{2N-1}$ is $\left( 2N-1\right) $, and that of $\mathcal{R}\simeq CP^{N-1}$
is $2\left( N-1\right) $.

In the kinematic approach to the geometric phase theory,  three kinds of curves $%
\mathcal{C}\subset \mathcal{B}$ of varying degrees of smoothness, and their
projections $C=\pi \left[ \mathcal{C}\right] $, are needed for specific
purposes. With monotonic parametrization, we write uniformly in all cases:%
\begin{equation}
\begin{array}{c}
\mathcal{C}=\left\{ \psi \left( s\right) \in \mathcal{B}|\text{ }s_{1}\leq
s\leq s_{2}\right\}\subset \mathcal{B} \overset{\pi }{\longrightarrow } \\ 
C=\pi \left[ \mathcal{C}\right] =\left\{ \rho \left( s\right) =\psi \left(
s\right) \psi ^{\dag }\left( s\right) \in \mathcal{R}|\text{ }s_{1}\leq
s\leq s_{2}\right\}\subset \mathcal{R} 
\end{array}
\label{NPC.2.2}
\end{equation}%
For geodesics we require $\mathcal{C}$ to be continuous twice-differentiable
with non-orthogonal endpoints. For \text{NPC}'s we need $\mathcal{C}$ continuous
once-differentiable with every pair of points on $\mathcal{C}$
non-orthogonal. Finally, for geometric phases to exist we need $\mathcal{C}$
continuous, piecewise once-differentiable with non-orthogonal endpoints. We
will find that we have the  inclusion relations:%
\begin{equation}
\text{Geodesics}\subset \mathrm{\text{NPC}'s}\subset \text{Curves  with geometric 
phase}  \label{NPC.2.3}
\end{equation}

Two non-orthogonal vectors $\psi ,\phi \in \mathcal{B}$ are defined to be
``in phase" in the Pancharatnam sense if:%
\begin{equation}
\left( \phi ,\psi \right) = \overline{\left( \psi ,\phi \right)}  >0  \label{NPC.2.4}
\end{equation}%
i.e., $\left( \phi ,\psi \right) $ is a positive real number.
More generally, the phase of $\psi $ with respect to $\phi $ is defined to
be $\arg \left( \phi ,\psi \right) $.

The lowest order Bargmann invariant (\text{BI}) involves three
pairwise non-orthogonal vectors $\psi _{1},\psi _{2},\psi _{3}\in \mathcal{B}
$ and is the expression (for $\dim \mathcal{H}\geq 2$):%
\begin{equation}
\begin{array}{c}
\Delta _{3}\left( \psi _{1},\psi _{2},\psi _{3}\right) =\left( \psi
_{1},\psi _{2}\right) \left( \psi _{2},\psi _{3}\right) \left( \psi
_{3},\psi _{1}\right) = \mathrm{Tr} \left( \rho _{1}\rho _{2}\rho _{3}\right)  \\ 
\rho _{j}=\psi _{j}\psi _{j}^{\dag }\in \mathcal{R},\text{ }j=1,2,3 .%
\end{array}
\label{NPC.2.5}
\end{equation}%
In a straightforward way this can be generalized to the $n$-th order \text{BI} $%
\Delta _{n}\left( \psi _{1},\psi _{2},...,\psi _{n}\right) $, provided
successive pairs of vectors are non-orthogonal.

The geometric phase for a curve $C\subset \mathcal{R}$ (of appropriate type)
is defined and most easily calculated using any lift $\mathcal{C}\subset 
\mathcal{B}$ of it, and it is the difference between a total (or Pancharatnam) phase
and a dynamical phase:%
\begin{equation}
\begin{array}{c}
\varphi _{g}\left[ C\right] =\varphi _{\text{tot}}\left[ \mathcal{C}\right]
-\varphi _{\text{dyn}}\left[ \mathcal{C}\right]  \\ 
\varphi _{\text{tot}}\left[ \mathcal{C}\right] =\arg \left( \psi \left(
s_{1}\right) ,\psi \left( s_{2}\right) \right)  \\ 
\varphi _{\text{dyn}}\left[ \mathcal{C}\right] =\text{Im}\int
_{s_{1}}^{s_{2}}ds\left( \psi \left( s\right) ,\psi'\left(
s\right) \right) 
\end{array}
\label{NPC.2.6}
\end{equation}

The original connection between \text{BI}'s and geometric phases involved the use
of geodesics in $\mathcal{R}$ and their lifts to $\mathcal{B}$. For any $%
C\subset \mathcal{R}$ (of appropriate type) its  length is defined as the
non-degenerate functional:%
\begin{equation}
L\left[ C\right] =\int_{s_{1}}^{s_{2}}ds\left\{ \left\Vert \frac{%
d\psi \left( s\right) }{ds}\right\Vert ^{2}-\left\vert \left( \psi \left(
s\right) ,\frac{d\psi \left( s\right) }{ds}\right) \right\vert ^{2}\right\}
^{1/2}  \label{NPC.2.7}
\end{equation}%
and the second-order ordinary differential equation determining geodesics
arises from here as the corresponding Euler-Lagrange equation. Solving it
one finds that given any two non-orthogonal points $\rho _{1},\rho _{2}\in 
\mathcal{R}$ and choosing $\psi _{1}\in \pi ^{-1}\left( \rho _{1}\right) ,$ $%
\psi _{2}\in \pi ^{-1}\left( \rho _{2}\right) $ in phase with one another in
the Pancharatnam sense, the (shortest) geodesic from $\rho _{1}$ to $\rho _{2}
$ possesses the following lift to $\mathcal{B}$:%
\begin{equation}
\psi \left( s\right) =\psi _{1}\cos s+\frac{\psi _{2}-\psi _{1}\left( \psi
_{1},\psi _{2}\right) }{\sqrt{1-\left( \psi _{1},\psi _{2}\right) ^{2}}}\sin
s;\text{ }0\leq s\leq \cos ^{-1}\left( \psi _{1},\psi _{2}\right) \in \left(
0,\pi /2\right)   \label{NPC.2.8}
\end{equation}%
We see that $\psi \left( s\right) $ is a real (positive) linear combination
of $\psi _{1}$ and $\psi _{2}$, and $\psi \left( s\right) ,\psi \left(
s^{\prime }\right) $ are in phase in the Pancharatnam sense for all $%
s,s^{\prime }$. Then the \text{BI}-geometric phase connection is:%
\begin{equation}
\arg \Delta _{3}\left( \psi _{1},\psi _{2},\psi _{3}\right) = 
-\varphi _{g}\left[ \text{\text{geodesic triangle in }}\mathcal{R}\text{ 
\text{with vertices }}\rho _{1},\rho _{2},\rho _{3}\right] 
\label{NPC.2.9}
\end{equation}%
(This easily generalizes to higher-order \text{BI}'s). Notice that while the
left-hand side depends only on the vertices, the definition of the
right-hand side requires that they be connected in some manner, here by
geodesics.

Now we come to the definition of a \text{NPC}.  A curve $C\subset \mathcal{R}$ (of
appropriate type), along with any lift $\mathcal{C}\subset \mathcal{B}$, is
a \text{NPC} if:%
\begin{equation}
\Delta _{3}\left( \psi \left( s\right) ,\psi \left( s^{\prime }\right) ,\psi
\left(s^{\prime \prime }\right)\right) =  
\overline{ \Delta _{3}\left( \psi \left( s\right) ,\psi \left( s^{\prime }\right) ,\psi
\left(s^{\prime \prime }\right) \right) } 
> 0. \quad  \forall s,s^{\prime },s^{\prime \prime }\in \left[ s_{1},s_{2}\right]  .
\label{NPC.2.10}
\end{equation}

From Eq.(\ref{NPC.2.8}) we see that every geodesic is a \text{NPC}, but it turns
out that for $\dim \mathcal{H}\geq 3$ the converse is not true. The key
property of a \text{NPC} is that:%
\begin{equation}
\varphi _{g}\left[ \text{\text{any connected portion of a NPC}}\right] =0
\label{NPC.2.11}
\end{equation}%
so connected portions of a \text{NPC} are themselves \text{NPC}'s. This definition is
designed just so that in place of the connection (\ref{NPC.2.9}) we have the
vastly extended relation :%
\begin{equation}
\begin{array}{c}
\arg \Delta _{3}\left( \psi _{1},\psi _{2},\psi _{3}\right) =-\varphi _{g}[%
\text{\text{ ``triangle" in }}\mathcal{R}\text{ \text{with vertices}} \\ 
\rho _{1},\rho _{2},\rho _{3}\text{ \text{joined pairwise by } NPC}\text{'s}%
]%
\end{array}
\label{NPC.2.12}
\end{equation}%
(This also generalizes to higher orders).  Hereafter it will be convenient to
denote by $N_{1,2}$ a \text{NPC} from $\rho _{1}$ to $\rho _{2}$ in $\mathcal{R}$, 
and by $\mathcal{N}_{1,2}$ a lift of it to $\mathcal{B}$.

At this point we bring in the basic differential-geometric objects which are
important for the following work. The dynamical phase $\varphi _{\text{dyn}}\left[ 
\mathcal{C}\right] $ in Eq.(\ref{NPC.2.6}) is the integral along $\mathcal{C}
$ \ of a one-form $A$ on $\mathcal{B}$:%
\begin{equation}
\varphi _{\text{dyn}}\left[ \mathcal{C}\right] =\int_{\mathcal{C}}A\text{ },%
\text{ \ }A=-i\psi ^{\dag }d\psi   \label{NPC.2.13}
\end{equation}%
This connection one-form is not the pull-back via $\pi ^{\ast }$ of any one-form on the
ray space $\mathcal{R}$.  However, the exterior derivative $dA$, its curvature,  is the pull-back of a closed non-degenerate (symplectic)
two-form $\omega$ on $\mathcal{R}$:
\begin{equation}
dA=\pi ^{\ast }\omega \text{ },\text{ }d\omega =0\text{ },\text{ }\omega 
\text{ \text{non-degenerate on }}\mathcal{R}  \label{NPC.2.14}
\end{equation}%
If $\mathcal{S}\subset \mathcal{B}$ is any smooth connected two-dimensional
surface with projection $S=\pi \left[ \mathcal{S}\right] \subset \mathcal{R}$, we have:%
\begin{equation}
\oint_{\partial \mathcal{S}}A = \int_{\mathcal{S}%
}dA=\int_{S}\omega   \label{NPC.2.15}
\end{equation}%
As a consequence, if in Eq.(\ref{NPC.2.6}) we take $C$ to be closed, and its
lift $\mathcal{C}$ to be also closed, we find that the geometric phase is a
symplectic area.  This is, if $\partial C= \emptyset$, $\partial \mathcal{C}= \emptyset$ and $S$ is any surface
such that $\partial S=C$, then:%
\begin{equation}
\varphi _{g}\left[ C\right] =-\varphi _{\text{dyn}}\left[ \mathcal{C}\right]
=-\oint_{\mathcal{C}}A=-\int_{S}\omega \, .
\label{NPC.2.16}
\end{equation}%
Explicit forms for $A$ and $\omega $ in local (Darboux) coordinates may be
easily obtained.

As mentioned earlier, it has been shown that the entire theory of the
geometric phase can be built up starting from \text{BI}'s and \text{NPC}'s. In this
process, the fact that (for $\dim \mathcal{H}\geq 3$) there are infinitely
many \text{NPC}'s connecting any two non-orthogonal points $\rho _{1},\rho
_{2}\in \mathcal{R}$, as against a single geodesic, has led to the concept
of \text{NPM}'s.  The precise definition of a  \text{NPM} will be given in the next
section.   At one extreme, a single  \text{NPC} is an example of a one-dimensional  \text{NPM}.
At the other extreme, for $\mathcal{H}$ of finite dimension, one can
ask for the maximum possible dimension of a  \text{NPM}.   It has been shown that a  
\text{NPM} must be an isotropic submanifold in $\mathcal{R}$, bringing in the
symplectic structure of $\mathcal{R}$.  However it has also been shown that
isotropy is not sufficient to obtain the  \text{NPM} property.  This  ``gap" will be
examined, and a complete characterization of  \text{NPM}'s obtained, in the next
section.

\section{NPM's and Isotropic Totally Geodesic Submanifolds.} \label{s:isotropic}

We begin by assembling a set of background results on geometric phases for
general curves in $\mathcal{R}$. As with the notations $N_{1,2}$ and $%
\mathcal{N}_{1,2}$ for  \text{NPC}'s, by $C_{1,2}$ we will mean a general curve
(of appropriate kind) connecting given $\rho _{1},\rho _{2}\in \mathcal{R}$,
and $\mathcal{C}_{1,2}$ a lift of it. The general non-additivity of
geometric phases is expressed by:%
\begin{equation}
\begin{array}{c}
\varphi _{g}\left[ C_{1,2}\cup C_{2,3}\cup ...\cup C_{n-1,n}\right] =\varphi
_{g}\left[ C_{1,2}\right] +\varphi _{g}\left[ C_{2,3}\right] +...+\varphi
_{g}\left[ C_{n-1,n}\right]  \\ 
-\arg \Delta _{n}\left( \psi _{1},\psi _{2},...,\psi _{n}\right) ;\text{ }%
\rho _{j}=\psi _{j}\psi _{j}^{\dag },\text{ }j=1,2,...,n%
\end{array}
\label{NPC.3.1}
\end{equation}%
An exception occurs for $n=3$ if we choose $\rho _{3}=\rho _{1}$. Then:%
\begin{equation}
\varphi _{g}\left[ C_{1,2}\cup C_{2,1}\right] =\varphi _{g}\left[ C_{1,2}%
\right] +\varphi _{g}\left[ C_{2,1}\right]   \label{NPC.3.2}
\end{equation}

For a curve $C_{1,2}$, let us denote by $\widetilde{C}_{1,2}$ the reversed
curve from $\rho _{2}$ to $\rho _{1}$; then the geometric phase changes
sign, and from Eq.(\ref{NPC.3.2}) we get for two curves from $\rho _{1}$ to 
$\rho _{2}$:%
\begin{equation}
\varphi _{g}\left[ C_{1,2}^{\prime }\right] =\varphi _{g}\left[ C_{1,2}%
\right] -\varphi _{g}\left[ C_{1,2}\cup \widetilde{C}_{1,2}^{\prime }\right] 
\label{NPC.3.3}
\end{equation}%
As the argument of the second term is a closed loop, we can use Eq.(\ref%
{NPC.2.16}) to get:%
\begin{equation}
\varphi _{g}\left[ C_{1,2}^{\prime }\right] =\varphi _{g}\left[ C_{1,2}%
\right] -\int_{S}\omega ,\text{ \ }\partial S=C_{1,2}\cup \widetilde{%
C}_{1,2}^{\prime }  \label{NPC.3.4}
\end{equation}%
This relation shows how the geometric phase changes if the endpoints are
kept fixed and the connecting curve is varied smoothly.

If in Eq.(\ref{NPC.3.2}) we take $C_{2,1}$ to be a  \text{NPC} $ N_{2,1}$ and
then use Eq.(\ref{NPC.2.16}), we get:%
\begin{equation}
\varphi _{g}\left[ C_{1,2}\right] =\varphi _{g}\left[ C_{1,2}\cup N_{2,1}%
\right] =\int_{S}\omega ,\text{ \ }\partial S=C_{1,2}\cup N_{2,1}
\label{NPC.3.5}
\end{equation}%
This is the most general way in which the geometric phase for an open curve
can be converted to that for a closed loop.

In order to set up the definition of a \text{NPC}, we recall how to obtain Eq.(%
\ref{NPC.2.11}) from Eq.(\ref{NPC.2.10}) for a single \text{NPC} . Given a \text{NPC}  $N
$, Eq.(\ref{NPC.2.10}) allows us to construct particular lifts $\mathcal{N}%
_{0}$ which have the global Pancharatnam property. For a fiducial $\rho
_{0}\in N$, we choose $\psi _{0} \in \pi ^{-1}\left( \rho _{0}\right) $.
Then for each $\rho \in N$, we choose \cite{MAEMM,man} $\psi = \rho \psi_0/ \sqrt{\mathrm{Tr}(\rho \rho_0)}$ 
and thus build
up $\mathcal{N}_{0}$. Eq.(\ref{NPC.2.10}) then shows that any two vectors $%
\psi ,\psi ^{\prime }\in \mathcal{N}_{0}$ are also in phase in the
Pancharatnam sense, so $\mathcal{N}_{0}$ is globally ``in phase". The
vanishing of \ geometric phases for all connected portions of $N$, Eq.(\ref%
{NPC.2.11}), is now immediate. In fact, both total and dynamical phases
vanish individually.

The definition of a \text{NPM}  can now be given in three equivalent ways. Let $M$
be a (regular) simply connected submanifold in $\mathcal{R}$, and write the
identification map as usual as: $i_{M}:M\hookrightarrow \mathcal{R}$. Then:%
\begin{equation}
\begin{array}{c}
M \text{ is a NPM} \\ 
\Leftrightarrow \text{\text{every} } C\subset M \text{ is a NPC} \\ 
\Leftrightarrow \Delta _{3}\left( \psi _{1},\psi _{2},\psi _{3}\right) = \overline{\Delta _{3}\left( \psi _{1},\psi _{2},\psi _{3}\right)} %
> 0\text{ }\forall \rho _{j}=\psi _{j}\psi _{j}^{\dag
}\in M,\text{ }j=1,2,3 \\ 
\Leftrightarrow \text{\text{there exist lifts }}\mathcal{M}_{0}\text{ 
\text{which are globally `` in phase"}}%
\end{array}
\label{NPC.3.6}
\end{equation}%
The third statement follows from the second by a construction similar to the 
\text{NPC} case described above. It is a simple consequence of Eqs.(\ref{NPC.3.6}%
) that:%
\begin{equation}
\rho _{1},\rho _{2}\in M\Longrightarrow \mathrm{Tr}\left( \rho _{1}\rho _{2}\right) >0
\label{NPC.3.7}
\end{equation}%
so a \text{NPM} does not contain mutually orthogonal points. The isotropy
property of $M$ also follows easily:%
\begin{equation}
\begin{array}{c}
C\subset M,\text{ }\partial C= \emptyset,\text{ }C\text{ \textit{a }}%
NPC\Longrightarrow  \\ 
\int_{S}\omega _{M}=0,\text{ } \forall S\subset M,\text{ with }\partial S=C,\text{ 
}\omega _{M}=i_{M}^{\ast }\omega \Longrightarrow \omega _{M}=0 
\end{array}
\label{NPC.3.8}
\end{equation}%
as there is complete freedom in the choice of the closed loop $C\subset M$.
Therefore a \text{NPM} is necessarily isotropic.

Now we consider the situation in the reverse direction. For a regular
submanifold $M\subset \mathcal{R}$, which
obeys the isotropy condition $i_{M}^{\ast }\omega =0$, what additional
properties are needed to conclude that $M$ is a \text{NPM}? Let us assume
hereafter that the $M$ under consideration always obeys Eq.(\ref{NPC.3.7}).
Let the curves $C_{1,2},C_{1,2}^{\prime }$ and the surface $S$ with $\partial S=C_{1,2}\cup \widetilde{
C}_{1,2}^{\prime }$ all be chosen to lie within $M$. Then, given $i_{M}^{\ast }\omega =0$, from Eq.(\ref%
{NPC.3.4}) we have:%
\begin{equation}
\varphi _{g}\left[ C_{1,2}^{\prime }\right] =\varphi _{g}\left[ C_{1,2}%
\right]   \label{NPC.3.9}
\end{equation}%
Therefore $\varphi _{g}\left[ C_{1,2}\right] $ is unchanged by continuous changes
of the curve which preserve its endpoints; that is, $\varphi _{g}\left[ C_{1,2}\right] $
depends only on $\partial C_{1,2}$. This falls short of showing that, for a closed loop $C\subset M$ is such that $\partial S = C$ for a surface $S\subset M$, 
$\varphi _{g}\left[ C\right] $ always vanishes.

If now it is the case that for every pair of points $\rho _{1},\rho _{2}\in M
$, the geodesic from $\rho _{1}$ to $\rho _{2}$ lies totally in $M$, then
in Eq.(\ref{NPC.3.9}) we can take $C_{1,2}^{\prime }$ to be this geodesic
and then conclude that \ $\varphi _{g}\left[ C_{1,2}\right] =0$. This would
mean that every $C$ is a \text{NPC}, and $M$ a \text{NPM}.

Actually it is clear that a weaker property of $M$ would suffice: if for \
every $\rho _{1},\rho _{2}\in M$ there is \textit{at least one} \text{NPC} $%
N_{1,2}\subset M$, then again by taking $C_{1,2}^{\prime }=N_{1,2}$ in Eq.(%
\ref{NPC.3.9}) we reach the desired conclusion: $\varphi _{g}\left[ C_{1,2}%
\right] =0$ and every $C$ is a \text{NPC}. Equally well we can take $N_{1,2}$ in
Eq.(\ref{NPC.3.5}) to be this \text{NPC}, and then also by isotropy we get the
desired result. However, it would be inappropriate to assume the existence
of some \text{NPC}'s in the process of proving that all $C$ are \text{NPC}'s.

A submanifold $M\subset \mathcal{R}$ (obeying Eq.(\ref{NPC.3.7})) with the
property that the geodesics connecting pairs of points in $M$ lie totally in 
$M$ is said to be \textit{totally geodesic} \cite{hel}. We have therefore shown that a
(regular, simply connected) isotropic totally geodesic submanifold $M\subset 
\mathcal{R}$ is definitely a \text{NPM}.  However the converse is not true for a
simple reason.  In an $M$ which is isotropic and totally geodesic (therefore
a \text{NPM}) we can choose any regular submanifold $M^{\prime }\subset M$ which
will certainly be isotropic as well as a \text{NPM}, but will in general not be a
totally geodesic submanifold. This gap which remains can be closed by the
following argument.

Let us collect the conclusions so far obtained:%
\begin{equation}
\begin{array}{c}
\left( a\right) \text{ }M \text{ is a NPM} \Rightarrow M\text{ 
\text{is isotropic}} \\ 
\left( b\right) \text{ }M\text{ \text{ simply connected, isotropic totally geodesic }}%
\Rightarrow \text{ }M \text{ is a NPM} \\ 
\left( c\right) \text{ }M^{\prime }\text{ \text{a simply connected regular submanifold in \
an isotropic totally geodesic}} \\ 
\text{\text{submanifold }}\Longrightarrow M^{\prime } \text{ is a NPM}
\end{array}
\label{NPC.3.10}
\end{equation}

We now show by construction that (\ref{NPC.3.10}-c) holds in the reverse
direction as well. Dropping primes:%
\begin{equation}
\begin{array}{c}
M\text{ \text{is a NPM}}\Longrightarrow M \text{ \text{ is a
regular submanifold in \ an}} \\ 
\text{\text{isotropic totally geodesic submanifold}}%
\end{array}
\label{NPC.3.11}
\end{equation}%
The construction is as follows. Given the \text{NPM}  \ $M\subset \mathcal{R}$, we
select one of its lifts $\mathcal{M}_{0}$ which has the Pancharatnam `` in
phase" property globally (cfr. Eq.(\ref{NPC.3.6})):%
\begin{equation}
\begin{array}{c}
M\subset \mathcal{R,}\text{NPM}\longrightarrow \mathcal{M}_{0}\subset 
\mathcal{B},\text{ }\pi \left[ \mathcal{M}_{0}\right] =M ; \\ 
\psi ,\psi ^{\prime }\in \mathcal{M}_{0}\Longrightarrow \left( \psi ,\psi
^{\prime }\right) =  \overline{\left( \psi ,\psi
^{\prime }\right)} >0%
\end{array}
\label{NPC.3.12}
\end{equation}%
We pass now from $\mathcal{M}_{0}$ to its non-negative real linear hull,
namely $\widetilde{\mathcal{M}}_{0}\subset \mathcal{B}$ made up of all
(normalized) real non-negative linear combinations of all sets of vectors in 
$\mathcal{M}_{0}$, hence $\widetilde{\mathcal{M}}_0$ is simply connected.   
Clearly $\mathcal{M}_{0}\subseteq \widetilde{\mathcal{M}}%
_{0}$, and $\widetilde{\mathcal{M}}_{0}$ retains the property of isotropy
since it is a \text{NPM}: because of Eq.(\ref{NPC.3.12}) and the method of
construction of $\widetilde{\mathcal{M}}_{0}$, all total and dynamical
phases vanish for curves in $\widetilde{\mathcal{M}}_{0}$. In particular,
the second line of (\ref{NPC.3.12}) remains valid for all pairs of vectors in $\widetilde{\mathcal{M}}_{0}$.
Now however 
$\widetilde{\mathcal{M}}_{0}$ (more precisely $\widetilde{M}=\pi \left[ 
\widetilde{\mathcal{M}}_{0}\right] $) is totally geodesic since the
construction in Eq.(\ref{NPC.2.8}) of geodesics is totally in the real
domain. This completes the proof of Eq.(\ref{NPC.3.11}). $\Box $\\

It should be clear that we need to resort to this construction or extension $%
M\rightarrow \widetilde{M}$ only if $M$ is not already totally geodesic.
Then it is also clear that the extension involved is minimal.

At this point we can answer the question raised at the end of Sect.\ref%
{s:Bargmann} concerning the maximum possible dimension of a \text{NPM}, assuming
the dimension $N$ of $\mathcal{H}$ is finite. From the isotropy property it
is clear that this maximum is $\left( N-1\right) $, one half of the real
dimension of the ray space $\mathcal{R}$. This follows from $\mathcal{R}$
being a symplectic manifold of dimension $2\left( N-1\right) $. Therefore a 
\text{NPM} \ $M$ of dimension $\left( N-1\right) $ is in fact a Lagrangian
submanifold in $\mathcal{R}$ (i.e., maximal isotropic), and it is necessarily already totally
geodesic, since there is no possible extension of $M$ to a larger isotropic
submanifold.

\section{Illustrative Examples} \label{se:examples}
We now consider some examples of \text{NPM}'s, to which for illustrative purposes
the construction of the previous Section can be applied.  Since a single \text{NPC},
being one-dimensional, is the simplest instance of a \text{NPM}, we begin with this case.

The definition of a \text{NPC} is given in Eq.$\left(\ref{NPC.2.10}\right)  $. A more
explicit description has been developed in Ref.$\left[  9\right]  $ and is as
follows. Let two distinct non-orthogonal points $\rho_{1},\rho_{2}%
\in\mathcal{R}$ obeying: $\mathrm{Tr}\left(  \rho_{1}\rho_{2}\right)  >0$ be given. Let
$N\subset\mathcal{R}$ be a \text{NPC}  from $\rho_{1}$ to $\rho_{2}:$%
\begin{equation}%
\begin{array}
[c]{c}%
N=\left\{  \rho\left(  s\right)  \in\mathcal{R}|\text{ }s_{1}\leq s\leq
s_{2},\text{ }\rho\left(  s_{1}\right)  =\rho_{1},\text{ }\rho\left(
s_{2}\right)  =\rho_{2}\right\}  \subset\mathcal{R}\\
\mathrm{Tr} \left(  \rho\left(  s\right)  \rho\left(  s^{\prime}\right)  \rho\left(
s^{\prime\prime}\right)  \right)  =\text{ \text{real positive }}\forall
s,s^{\prime},s^{\prime\prime}\in\left[  s_{1},s_{2}\right]
\end{array}
\label{sec4.1}%
\end{equation}

Choose vectors $\psi_{1},\psi_{2}\in\mathcal{B}$ projecting onto $\rho
_{1},\rho_{2}$ respectively, with $\left(  \psi_{1},\psi_{2}\right)  $ real
positive, so that \ $\psi_{1}$ and $\psi_{2}$ are in phase in the Pancharatnam
sense. As shown in the previous Section, we can construct a lift
$\mathcal{N}_{0}$ of $N$ from $\psi_{1}$ to \ $\psi_{2}$ which has the global
Pancharatnam property:%
\begin{equation}%
\begin{array}
[c]{c}%
\mathcal{N}_{0}=\left\{  \psi_{0}\left(  s\right)  \in\mathcal{B}|\text{
}s_{1}\leq s\leq s_{2},\text{ }\psi_{0}\left(  s_{1}\right)  =\psi_{1},\text{
}\psi_{0}\left(  s_{2}\right)  =\psi_{2};\text{ }\rho\left(  s\right)
=\pi\left(  \psi_{0}\left(  s\right)  \right)  \right\} \subset    \mathcal{B}  \\
\left(  \psi_{0}\left(  s\right)  ,\psi_{0}\left(  s^{\prime}\right)  \right)
=\text{ \text{real positive }}\forall s,s^{\prime}\in\left[  s_{1}%
,s_{2}\right]
\end{array}
\label{sec4.2}%
\end{equation}

We express the endpoints of \ $\mathcal{N}_{0}$ as:%
\begin{equation}%
\begin{array}
[c]{c}%
\psi_{1}=e_{1},\text{ }\psi_{2}=e_{1}\cos\theta_{0}+e_{2}\sin\theta_{0},\text{
}\theta_{0}\in(0,\pi\left.  {}\right) \\
\left(  e_{i},e_{j}\right)  =\delta_{ij},\text{ }i,j=1,2
\end{array}
\label{sec4.3}%
\end{equation}

Denote by $\mathcal{H}_{\bot}$ the orthogonal complement in $\mathcal{H}$ to
the two-dimensional subspace spanned by $e_{1}$ and $e_{2}$:%
\begin{equation}
\mathcal{H}_{\bot}=\left\{  \phi\in\mathcal{H}|\text{ }\left(  e_{1}%
,\phi\right)  =\left(  e_{2},\phi\right)  =0\right\}  \label{sec4.4}%
\end{equation}
Then the vectors $\psi_{0}\left(  s\right)  \in\mathcal{N}_{0}$ can be
expressed as:%
\begin{equation}%
\begin{array}
[c]{c}%
\psi_{0}\left(  s\right)  =x_{1}\left(  s\right)  e_{1}+x_{2}\left(  s\right)
e_{2}+\chi\left(  s\right) \\
\chi\left(  s\right)  \in\mathcal{H}_{\bot}\\
\left\vert x_{1}\left(  s\right)  \right\vert ^{2}+\left\vert x_{2}\left(
s\right)  \right\vert ^{2}+\left(  \chi\left(  s\right)  ,\chi\left(
s\right)  \right)  =1
\end{array}
\label{sec4.5}%
\end{equation}
At $s=s_{1},s_{2}$ we have:%
\begin{equation}%
\begin{array}
[c]{c}%
x_{1}\left(  s_{1}\right)  =1,\text{ }x_{2}\left(  s_{1}\right)  =0,\text{
}\chi\left(  s_{1}\right)  =0\\
x_{1}\left(  s_{2}\right)  =\cos\theta_{0},\text{ }x_{2}\left(  s_{2}\right)
=\sin\theta_{0},\text{ }\chi\left(  s_{2}\right)  =0
\end{array}
\label{sec4.6}%
\end{equation}

If we set $s^{\prime}=s_{1},s_{2}$ in the positivity condition of
Eq.(\ref{sec4.2}) we find:%
\begin{equation}
x_{1}\left(  s\right)  ,\text{ }x_{1}\left(  s\right)  \cos\theta_{0}%
+x_{2}\left(  s\right)  \sin\theta_{0}\text{ \text{real positive }}\forall
s\in\left[  s_{1},s_{2}\right]  \label{sec4.7}%
\end{equation}
We may therefore replace $x_{1}\left(  s\right)  $ and $x_{2}\left(  s\right)
$, which are both real, by the expressions:%
\begin{equation}
x_{1}\left(  s\right)  =\sigma\left(  s\right)  \cos\theta\left(  s\right)
,\text{ }x_{2}\left(  s\right)  =\sigma\left(  s\right)  \sin\theta\left(
s\right)  \label{sec4.8}%
\end{equation}
subject to:%
\begin{equation}%
\begin{array}
[c]{c}%
0<\sigma\left(  s\right)  \leq1,\text{ }-\frac{\pi}{2}+\theta_{0}%
<\theta\left(  s\right)  <\frac{\pi}{2}\\
\theta\left(  s_{1}\right)  =0,\text{ }\theta\left(  s_{2}\right)  =\theta
_{0},\text{ }\sigma\left(  s_{1}\right)  =\sigma\left(  s_{2}\right)  =1
\end{array}
\label{sec4.9}%
\end{equation}

Of course, for a particular \text{NPC} \ these ranges may not be fully utilized.
For the squared norm of $\chi\left(  s\right)  $ we have:%
\begin{equation}
\left\Vert \chi\left(  s\right)  \right\Vert ^{2}=\left(  \chi\left(
s\right)  ,\chi\left(  s\right)  \right)  =1-\sigma\left(  s\right)  ^{2}\geq0
\label{sec4.10}%
\end{equation}
The remaining content of the positivity condition in Eq.(\ref{sec4.2}) is:%
\begin{equation}
\sigma\left(  s\right)  \sigma\left(  s^{\prime}\right)  \cos\left(
\theta\left(  s^{\prime}\right)  -\theta\left(  s\right)  \right)  +\left(
\chi\left(  s^{\prime}\right)  ,\chi\left(  s\right)  \right)  =\text{
\text{real positive }}\forall s^{\prime},s\in\left(  s_{1},s_{2}\right)
\label{sec4.11}%
\end{equation}
This leads to $\left(  \chi\left(  s^{\prime}\right)  ,\chi\left(  s\right)
\right)  $ being real. It can be seen quite easily that as a consequence it
should be possible to choose an orthonormal basis $\left(  e_{3}%
,e_{4},...\right)  $ for \ $\mathcal{H}_{\bot}$ such that:%
\begin{equation}%
\begin{array}
[c]{c}%
\chi\left(  s\right)  =%
{\displaystyle\sum\limits_{r=3,4,...}}
x_{r}\left(  s\right)  e_{r},\text{ }x_{r}\left(  s\right)  \text{
\text{real}}\\
\left\Vert \chi\left(  s\right)  \right\Vert ^{2}=%
{\displaystyle\sum\limits_{r=3,4,...}}
x_{r}\left(  s\right)  ^{2}=1-\sigma\left(  s\right)  ^{2}\in\left[
0,1\right)
\end{array}
\label{sec4.12}%
\end{equation}
Then $\left(  e_{1},e_{2},e_{3},...\right)  $ is an orthonormal basis for
$\mathcal{H}$, with the choice of $e_{3},e_{4},...$  depending in general on the
particular \text{NPC} $N$ and lift $\mathcal{N}_{0}$ being considered.

Summarizing, the vectors along the special lift $\mathcal{N}_{0}$ of $N$ are
real linear combinations of the basis vectors $\left(  e_{1},e_{2}%
,e_{3},...\right)  $:%
\begin{equation}
\psi_{0}\left(  s\right)  =%
{\displaystyle\sum\limits_{r=1,2,...}}
x_{r}\left(  s\right)  e_{r}=\sigma\left(  s\right)  \cos\theta\left(
s\right)  e_{1}+\sigma\left(  s\right)  \sin\theta\left(  s\right)  e_{2}%
+\chi\left(  s\right)  \label{sec4.13}%
\end{equation}
subject to the conditions in Eqs.(\ref{sec4.6}), (\ref{sec4.9}) and
(\ref{sec4.11}).   While the conditions at $s_{1}$ and $s_{2}$ are easy to state
and ensure, the non-local condition (\ref{sec4.11}) has the geometrical
meaning that for all $s^{\prime},s\in\left(  s_{1},s_{2}\right)  $ the real
unit vectors $\widehat{x}\left(  s^{\prime}\right)  =\left\{  x_{1}\left(
s^{\prime}\right)  ,x_{2}\left(  s^{\prime}\right)  ,x_{3}\left(  s^{\prime
}\right)  ,...\right\}  $ and \ $\widehat{x}\left(  s\right)  =\left\{
x_{1}\left(  s\right)  ,x_{2}\left(  s\right)  ,x_{3}\left(  s\right)
,...\right\}  $ must make an angle less than $\pi/2$ with each other.

Based on this description of the most general \text{NPC}  from $\rho_{1}$ to
$\rho_{2}$, a relatively simple class of \text{NPC} 's suggests itself. We extend
the pair $\left\{  e_{1},e_{2}\right\}  $ to an orthonormal basis $\left\{
e_{1},e_{2},e_{3},...\right\}  $ in $\mathcal{H}$ in any way we wish, and choose some
$m\in\left\{  3,4,...\right\}  $. Then $\left\{  e_{1},e_{2},...,e_{m}%
\right\}  $ is an orthonormal set in $\mathcal{H}$, and we limit ourselves to
vectors $\psi_{0}\left(  s\right)  \in \text{Sp}\left\{  e_{1},e_{2},...,e_{m}%
\right\}  $. Let $\mathbb{S}^{m-1}\subset\mathbb{R}^{m}$ be the real unit sphere in an
$m$-dimensional real Euclidean space. Within $\mathbb{S}^{m-1}$, let us choose the
region $\mathbb{S}_{+}^{m-1}$ where all $m$ coordinates are positive:%
\begin{equation}
\mathbb{S}_{+}^{m-1}=\left\{  \widehat{x}=\left(  x_{1},x_{2},...,x_{m}\right)  \in
\mathbb{S}^{m-1}|\text{ }x_{j}>0,\text{ }j=1,2,...,m\right\}  \subset \mathbb{S}^{m-1}
\label{sec4.14}%
\end{equation}

Then by choosing once-differentiable $\widehat{x}\left(  s\right)  \in
\mathbb{S}_{+}^{m-1}$ for $s_{1}<s<s_{2}$, with $\widehat{x}\left(  s_{1}\right)
=\left(  1,0,...\right)  $ and $\widehat{x}\left(  s_{2}\right)  =\left(
\cos\theta_{0},\sin\theta_{0},0,...\right)  $, we generate a \text{NPC} 
$\mathcal{N}_{0}\subset\mathcal{B}$ from $\psi_{1}$ to $\psi_{2}$ as follows:%
\begin{equation}%
\begin{array}
[c]{c}%
\psi_{0}\left(  s\right)  =%
{\displaystyle\sum\limits_{r=1.}^{m}}
x_{r}\left(  s\right)  e_{r}\\
\psi_{0}\left(  s_{1}\right)  =\psi_{1}=e_{1},\text{ }\psi_{0}\left(
s_{2}\right)  =\psi_{2}=e_{1}\cos\theta_{0}+e_{2}\sin\theta_{0}%
\end{array}
\label{sec4.15}%
\end{equation}

By construction we have ensured the \text{NPC}  condition:%
\begin{equation}
\left(  \psi_{0}\left(  s\right)  ,\psi_{0}\left(  s^{\prime}\right)  \right)
=\widehat{x}\left(  s\right)  \mathbf{\cdot}\, \widehat{x}\left(  s^{\prime
}\right)  >0, \quad \forall s,s^{\prime}\in\left[  s_{1},s_{2}\right].
\label{sec4.16}%
\end{equation}

As depicted in the figure, this \text{NPC} can be pictured as a
once-differentiable curve lying in $\mathbb{S}_{+}^{m-1}$ and running from $\left(
1,0,...,0\right)  $ to $\left(  \cos\theta_{0},\sin\theta_{0},0,...,0\right)
:$

\begin{figure}[h]
\begin{center}
\includegraphics[scale=0.7]{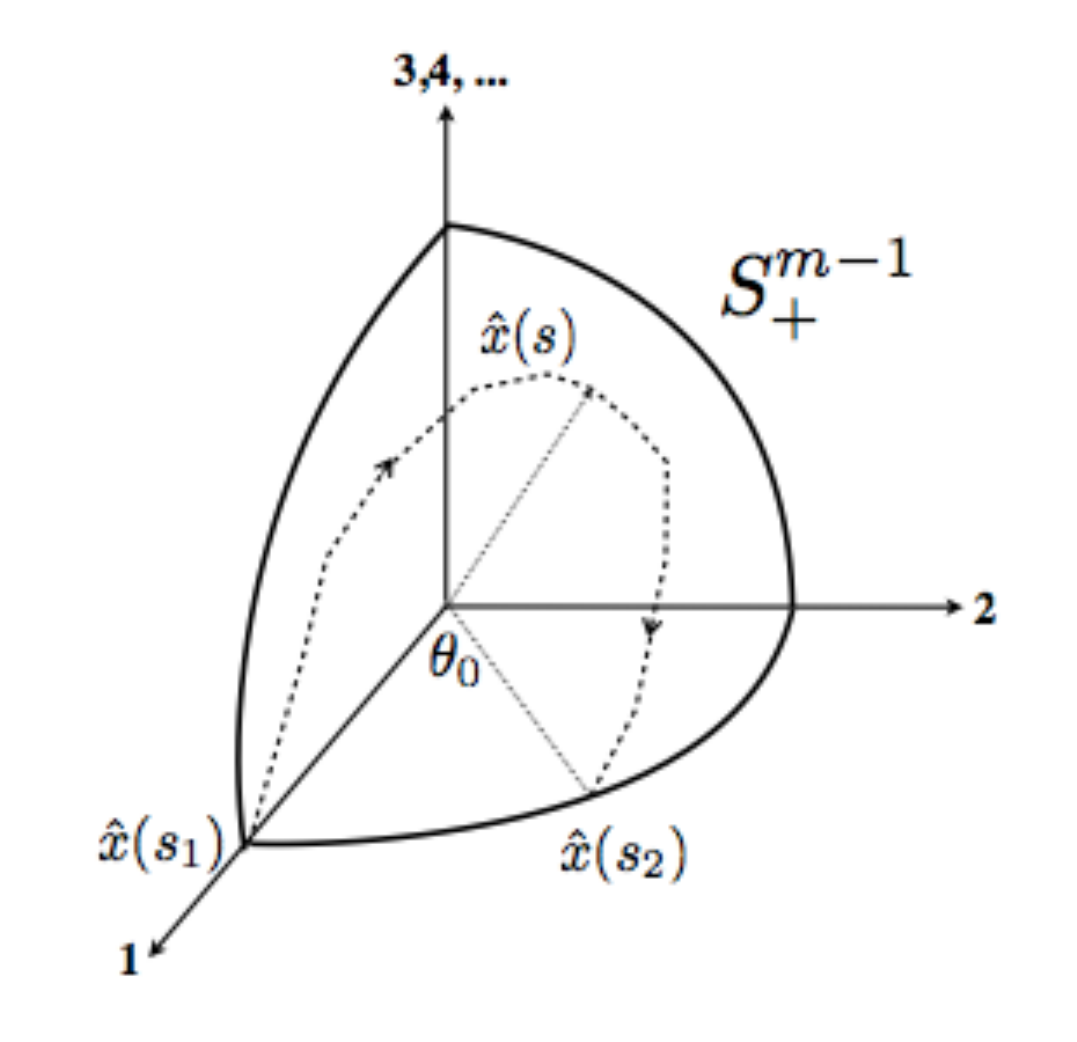}
\end{center}
\caption{The dotted curve represents a special class of $\text{NPC}$'s pictured on $\mathbb{S}_+^{m-1}$.}
\label{fig:mix_parameters}
\end{figure}

The passage from the one-dimensional \text{NPC}  \ $\mathcal{N}_{0}=\left\{
\psi_{0}\left(  s\right)  \right\}  \subset\mathcal{B}$ to its real
non-negative linear hull, in the manner of the previous section, leads to a
(generally higher-dimensional) submanifold $\widetilde{\mathcal{M}}_{0}%
\subset\mathcal{B}$. This construction can be carried out, for instance, by
forming all convex linear combinations of all subsets of vectors on
$\mathcal{N}_{0}$, and then normalizing the result. In $\mathcal{B}$, and
in the image in $\mathbb{S}_{+}^{m-1}$, we have:
\begin{equation}%
\begin{array}
[c]{c}%
\psi=c
{\displaystyle\sum\limits_{j}}
p_{j}\psi_{0}\left(  s_{j}\right)  ,\text{ }p_{j}>0,\text{ }%
{\displaystyle\sum\limits_{j}}
p_{j}=1,\text{ }\left\Vert \psi\right\Vert =1\\
\widehat{x}=c
{\displaystyle\sum\limits_{j}}
p_{j}\widehat{x}\left(  s_{j}\right)  ,\text{ }\widehat{x}\mathbf{\cdot
}\widehat{x}=1\text{\ }%
\end{array}
\label{sec4.17}%
\end{equation}

The image of \ $\widetilde{\mathcal{M}}_{0}$ on $\mathbb{S}_{+}^{m-1}$ is that it is
the minimal convex cone containing (the image of) $\mathcal{N}_{0}$. We can
see that the arc in the $1-2$ plane from $\widehat{x}\left(  s_{1}\right)  $
to $\widehat{x}\left(  s_{2}\right)  $ is included. Going back to
$\widetilde{\mathcal{M}}_{0}\subset\mathcal{B}$ and its image $\widetilde
{M}=\pi\left(  \widetilde{\mathcal{M}}_{0}\right)  \subset\mathcal{R}$, it is
clear that both isotropy and the totally geodesic property have been achieved
in a minimal manner starting from $\mathcal{N}_{0}$.

To deal with the most general \text{NPC}  (from $\psi_{1}$ to $\psi_{2}$) as
described in Eq.(\ref{sec4.13}) subject to Eqs.(\ref{sec4.7}), (\ref{sec4.9})
and (\ref{sec4.11} ) (and with the limitation to $\mathrm{span}\left\{  e_{1},e_{2}%
,e_{3},...,e_{m}\right\}  $), we must permit the choice of $e_{3},e_{4},...,e_{m}$
to depend on the particular \text{NPC}.  Then we see that in the figure above the
path of the real unit vector \ $\widehat{x}\left(  s\right)  $ can explore
regions of $\mathbb{S}^{m-1}$ outside of \ $\mathbb{S}_{+}^{m-1}$, while obeying the non-local
positivity condition in Eq.(\ref{sec4.2}). Thus for any $s$ and $s^{\prime}$,
the angle between $\widehat{x}\left(  s\right)  $ and $\widehat{x}\left(
s^{\prime}\right)  $ must be less than $\pi/2$. The component $x_{1}\left(
s\right)  >0$ throughout, while $x_{2}\left(  s\right)  ,x_{3}\left(
s\right)  ,...,x_{m}\left(  s\right)  $ can each be sometimes negative.
However, the image of \ $\widetilde{\mathcal{M}}_{0}$ is still the minimal
convex cone on $S^{m-1}$ containing the image of $\mathcal{N}_{0}$.

Turning to examples of \text{NPM}'s $M\subset\mathcal{R}$ of higher dimensions, we
consider two cases from Ref.$\left[  9\right]  $. The first one, in the
\ $\mathbb{S}_{+}^{m-1}$ picture just used to discuss single \text{NPC}'s, is to take (the
global Pancharatnam lift) $\mathcal{M}_{0}$ to be essentially $\ \mathbb{S}_{+}^{m-1}$:%
\begin{equation}%
\begin{array}
[c]{c}%
\mathcal{M}_{0}=\left\{  \psi\left(  \widehat{x}\right)  =%
{\displaystyle\sum\limits_{r=1}^{m}}
x_{r}e_{r}|\text{ }\widehat{x}\in \mathbb{S}_{+}^{m-1}\right\}  \subset\mathcal{B}\\
M=\pi\left(  \mathcal{M}_{0}\right)  \subset\mathcal{R}%
\end{array}
\label{sec4.18}%
\end{equation}

Since:%
\begin{equation}
\left(  \psi\left(  \widehat{x}\right)  ,\psi\left(  \widehat{x}^{\prime
}\right)  \right)  =\widehat{x}\mathbf{\cdot}\widehat{x}^{\prime}=\text{
\text{real }}>0 \label{sec4.19}%
\end{equation}
we have the \text{NPM} property for $M$:%
\begin{equation}
\Delta_{3}\left(  \psi\left(  \widehat{x}\right)  \psi\left(  \widehat
{x}^{\prime}\right)  \psi\left(  \widehat{x}^{\prime\prime}\right)  \right)
=  (\widehat{x}\mathbf{\cdot}\widehat{x}^{\prime})(\widehat{x}^{\prime
}\mathbf{\cdot}\widehat{x}^{\prime\prime})(\widehat{x}^{\prime\prime
}\mathbf{\cdot}\widehat{x} )= \text{ \ \text{real }} >0 \label{sec4.20}%
\end{equation}

In this case, as is also obvious from the definition of $\mathcal{M}_{0}$, its
real non-negative linear hull is itself: $\widetilde{\mathcal{M}}%
_{0}=\mathcal{M}_{0}$, so $M$ is already both isotropic and totally geodesic.

The second more concrete example involves a set of real Schr\"{o}dinger wave
functions in $\mathcal{H}=L^{2}\left(  \mathbb{R}^{N}\right)  $. We start from
the ground-state wave function of the $N$-dimensional isotropic simple
harmonic oscillator and and all its spatial translates:
\begin{equation}
\begin{array}{c}
\psi_{0}\left(  \mathbf{x}\right)  =\pi^{-N/4}\exp\left(  -
\mathbf{x\cdot x}/2\right)  ,\text{ }\mathbf{x\cdot x}=%
{\displaystyle\sum\limits_{j=1}^{N}}x_j^2\\
\psi_{\mathbf{y}}(x)=\psi_0(\mathbf{x-y}), \; \mathbf{y}\in\mathbb{R}^N
\end{array}
\label{sec4.21}%
\end{equation}
All these wave functions are normalized and pointwise real positive, and taken
together they define $\mathcal{M}_{0}$:%
\begin{equation}
\mathcal{M}_{0}=\left\{  \psi_{\mathbf{y}}\left(  \mathbf{x}\right)  |\text{
}\mathbf{y}\in\mathbb{R}^{N}\right\} \subset \mathcal{B} \subset \mathcal{H}  \label{sec4.22}%
\end{equation}
As all inner products $\left(  \psi_{\mathbf{y}},\psi_{\mathbf{y}^{\prime}%
}\right)  $ are trivially real positive, $M=\pi\left(  \mathcal{M}_{0}\right)
$ is clearly an $N$-dimensional \text{NPM}  in $\mathcal{R}$. However, on its own,
$M$ is not totally geodesic. The extension of $\mathcal{M}_{0}$ to its real
non-negative linear hull can be accomplished by first constructing ``convex
combinations" of the wave functions $\psi_{\mathbf{y}}\left(  \mathbf{x}%
\right)  $, namely:%
\begin{equation}
\psi\left(  \mathbf{x}\right)  = c \int   p\left(  \mathbf{y}\right)
\exp [ - \left(  \mathbf{x}-\mathbf{y}\right)
\mathbf{\cdot}\left(  \mathbf{x}-\mathbf{y}\right)/2] \, \, d^{N}y ,\qquad
p\left(  \mathbf{y}\right)   \geq0 \label{sec4.23}%
\end{equation}
and then fixing $c$ so that $\psi\left(  \mathbf{x}\right)  $ is normalized
(here we must permit choices of $p\left(  \mathbf{y}\right)  $ involving Dirac
delta functions as well).
This process clearly involves a genuine (minimal) enlargement of
$\mathcal{M}_{0}$ to $\widetilde{\mathcal{M}}_{0}$, and then the totally
geodesic property as well as isotropy is achieved for $\widetilde{M}%
=\pi\left(  \widetilde{\mathcal{M}}_{0}\right)  $.

The same reasoning may be applied to the class of Generalized Gaussian states \cite{SSM} of the kind:
\begin{equation}
\psi_{\mathbf{y},U}\left(  \mathbf{x}\right)  =\pi^{-N/4} (\det U)^{1/4} 
\exp\left(  -
\mathbf{(x-y)}\cdot U\mathbf{(x-y)}/2\right) \label{sec4.21bis}%
\end{equation}
where $\mathbf{y}$ is again an $N$-dimensional translation vector while $U$ is a real positive definite symmetric matrix.
In this case $M=\pi\left(  \mathcal{M}_{0}\right)$ is an $N+N(N+1)/2$-dimensional \text{NPM} and the analogue of formula (\ref{sec4.23}) involves also an integral over the $N(N+1)/2$ 
variables that parametrize the space of real positive definite symmetric matrices, hence yielding a quadratic increase in the dimension of the manifold.

\section{Concluding Remarks}

It is well appreciated that the concept of the geometric phase belongs to the
basic foundations of quantum mechanics. Its study has progressively
revealed many important aspects of the mathematical structure of the subject
and the interrelations among them. The introduction of the concepts of
Bargmann invariants and null phase curves has added considerable richness to
the subject.

On the other hand, the unravelling of the basic geometric features of the
state or ray spaces of quantum mechanics has been receiving considerable
attention \cite{EMM}. It is quite remarkable that these spaces are simultaneously
manifolds with Riemannian metric structures and symplectic structures. The work
in this paper has brought these two aspects very close together in the
context of the geometric Phase, by showing that null phase manifolds can be
fully characterized only by combining these structures suitably. At an
elementary level, a null phase manifold in ray space is a submanifold in which all
``evolutions" have identically vanishing geometric phases. However, the fact
that its understanding needs both the metric and symplectic structures of ray
space \ is quite remarkable, and can be expected to shed more light on the
foundations of quantum mechanics.

\section*{Acnowldegments}
G.M. would like to acknowledge  the support provided by the Santander/UCIIIM University Chair of Exccellence Programme 2011-12.

\end{document}